\documentclass[review]{elsarticle}

\usepackage{lineno,hyperref}
\usepackage{color}

\begin{document}

\begin{frontmatter}

\title{Agnostic conservative down-sampling for optimizing statistical representations and PIC simulations}

\author[GU,UNN,IAP]{A.~Gonoskov}

\address[GU]{Department of Physics, University of Gothenburg, SE-41296 Gothenburg, Sweden}
\address[UNN]{Lobachevsky State University of Nizhni Novgorod, Nizhny Novgorod 603950, Russia}
\address[IAP]{Institute of Applied Physics, Russian Academy of Sciences, Nizhny Novgorod 603950, Russia}

\date{\today}

\begin{abstract}
In particle-in-cell simulations and some other statistical computations, the representation of modelled distributions with tracked macro-particles can become locally excessive. Merging or resampling dense clusters or highly-populated phase space volumes may, however, remove or affect small-scale peculiarities in the modelled distribution or cause local changes of conserved quantities, such as energy and momenta. This may lead to additional noise, reduced accuracy or even unphysical effects. Here we describe a probabilistic algorithm for reducing the number of macro-particles in such clusters or volumes so that all the distribution functions are not affected on average and an arbitrary number of conservation laws, distribution central moments and contributions to the grid quantities (such as charge and current density) are preserved.
\end{abstract}

\begin{keyword}
particle-in-cell \sep particle merging \sep particle coalescing \sep resampling
\end{keyword}

\end{frontmatter}


\section{Introduction}

In particle-in-cell (PIC) simulations and some other statistical computations, where the modelled distribution is represented by an ensemble of tracked macro-particles, it may be beneficial or even necessary to reduce the number of macro-particles, increasing the weight of their contribution. This procedure, also referred to as down-sampling, merging or coalescing, is typically applied to the phase space regions that are overpopulated with macro-particles. The procedure can be repeatedly performed to allow for better representation of the other regions (where up-sampling is performed) or to mitigate the growth of macro-particle number, when modeling emission from boundaries, ionization or the generation of particles and photons. Certainly, the algorithm of down-sampling must be non-destructive for the modelled physics of interest to a largest possible degree. Several authors have performed a thorough consideration of various aspects and proposed efficient algorithms.

One of the main concerns is attributed to the consequences of not preserving intrinsically conserved quantities, such as the charge, energy or momentum of the down-sampled subset of macro-particles. Although the random variations may cancel each other in the long-term range, the changes of such quantities may also add up and cause artificial effects. For example, local variation of charge density can lead to the rise of global electric field and/or to an artificial heating. In addition, even minor charge relocation that is not supplemented with corresponding currents show up as the noise in the electromagnetic field (unless the field solver includes the corrections by the solution of the Poisson's equation). Lapenta and Brackbill proposed algorithms for coalescing two macro-particles into one, and also splitting one macro-particle into two, that preserve charge assignment at grid points \cite{lapenta.jcp.1994}. The same authors have then proposed an algorithm to perform replacement of all macro-particles within a given cell with a new set of macro-particles, preserving the contributions to the grid and also maximizing the uniformity of physical quantities among the new macro-particles \cite{lapenta.cpc.1995}. Assous et al. have proposed an algorithm for such replacement so that it preserves mesh charge and current densities, and also the energy of macro-particles \cite{assous.jcp.2003}. This approach has been further developed in Ref.~\cite{welch.jcp.2007}.

Recently, the problem of arranging down-sampling has gained a significant interest as a necessary component for PIC simulations of cascaded electron-positron pair production due to the effect of strong-field quantum electrodynamics in strong laser fields \cite{nerush.prl.2011, elkina.prstab.2011, sokolov.pop.2011, ridgers.jcp.2014, gonoskov.pre.2015, chang.2017}. The simulations of experimental proposals for the upcoming and future high-intensity laser facilities have shown the prolific growth of particle and photon number by many orders of magnitude \cite{gonoskov.prx.2017, tamburini.scirep.2017, vranic.scirep.2018, efimenko.scirep.2018, efimenko.pre.2019}. Such simulations thus require rapid down-sampling, whereas any inaccuracies of this procedure may grow up with the cascade development. Timokhin \cite{timokhin.mnras.2010} used a global procedure that repeatedly removes a random macro-particle and redistribute its weight among the others of the same kind. A similar procedure with the redistribution of also charge and energy has been used by Nerush et al. \cite{nerush.prl.2011}. {In case of using coalescing/merging, even if not all the conservation laws can be preserved, the introduced distortions can be mitigated by the closeness of merged macro-particles in the phase space. Rjasanow et al. demonstrated rigorously that the locality of the merged subsets of macro-particles is central for reducing the introduced errors and proposed using clustering via hierarchical binary space subdivision \cite{rjasanow.jcp.1996, rjasanow.jcp.1998}. Martin and Cambier considered clustering based on octree binning in velocity space \cite{martin.jcp.2016}.} Luu et al. proposed to select {dense} clusters based on the concept of the Voronoi diagram \cite{luu.cpc.2015}. {Another aspect is that merging a subset} into a single macro-particle {does not provide enough degrees of freedom} to conserve both energy and momentum. Vranic et al. {considered} merging into a pair of macro-particles with momenta chosen so that both energy and momentum are preserved simultaneously \cite{vranic.cpc.2015} {(merging subsets into a pair of macro-particles had been earlier described and analyzed in Ref.~\cite{rjasanow.jcp.1996, rjasanow.jcp.1998})}. Pfeiffer et al. proposed a statistical method for massive resampling that conserves momentum and energy \cite{pfeiffer.cpc.2015}. Faghihi et al. reported on the development of an algorithm for massive down-sampling that preserves any number of particle and grid quantities \cite{faghihi.arxiv.2017}.

In this paper we concern one more aspect of ensuring that down-sampling {does not affect significantly the physical process being simulated.} We discuss how a down-sampling procedure may affect small-scale {details} in the particle distribution functions and describe an algorithm that, apart from {the capability to preserve several physical quantities, tend to preserve} all the peculiarities independently of their scales. In this context, we refer to this algorithm as an \textit{agnostic} algorithm because it does not assume that the down-sampled subset of macro-particles represents a uniform part of distribution (see rigorous definition in Sec.~\ref{sec:agnostic}). {The method is based on probabilistic redistribution of weights within a local cluster and does not interfere with the way of clustering. In the simplest cases, it can be applied to a subset of neighboring macro-particles (1D case) or to macro-particles within densely populated cells, which can be naturally facilitated by the corresponding strategy of storing particles in PIC codes (see example in the end of the paper). To achieve better accuracy one can arrange applying the method to the subsets formed by some clustering method.}

{To demonstrate the declared capabilities of the algorithm we apply the algorithm being configured to preserve one or several first central moments of 1D particle distribution within subsets of neighboring macro-particles selected from a 1D distribution. By doing so we achieve the expected result: the increase of the number of preserved first central moments reduces the variance of the density computed through weighing over introduced cells. A sharp jump in the initial density is used to demonstrate that the improvement is achieved without causing any systematic deviations at any scale, which demonstrates and illustrates the effect of down-sampling being agnostic. To demonstrate explicitly the capability to preserve several quantities simultaneously we show how the algorithm can be configured to preserve grid values of density calculated with the cloud-in-cell weighing in 2D case. Note that these examples give just a generic confirmation that the algorithm can be implemented in the proposed form and have the declared properties, whereas the benefits and aspects of using the algorithm in realistic conditions is a matter of more context-based analysis. Some results can be found in Ref.~\cite{muraviev.arxiv.2020} where the analysis is carried out using several pertinent physical problems, dedicated efficiency criteria, and based on a comparison with the results of other down-sampling methods. The algorithm has been made publicly available as an open-source tool within the hi-$\chi$ framework \cite{hi-chi}.}

\section{Agnostic down-sampling}
\label{sec:agnostic}

In many cases down-sampling implies the selection of macro-particles that are closely placed in the phase space and thus can either be merged into one or two macro-particles or replaced by a smaller set of new macro-particles. When doing so, we implicitly assume that the chosen macro-particles represent a uniform element of modelled distribution in the phase space. Thus, for example, merging of macro-particles may lead to the unification of two close but distinctly different peaks {in the distribution of particles when modeling} two spatially overlaid streams of particles {that propagate in two close directions}. One can probably avoid this by restricting the selection of particles to phase space volumes that are sufficiently small for the physics in question. This, however, may restrict the applicability by requiring higher density of macro-particles in the phase space for the selection to be possible. In addition, this requires some prior knowledge about the minimal scales of the modeled physical processes. Note that, although the spatial resolution is naturally limited by the grid step, the momentum resolution is not limited in PIC method. Let us therefore consider an alternative approach that is applicable without such prior knowledge and thus called here \textit{agnostic} down-sampling.

Firstly, to avoid adding any macro-particles to potentially empty regions of the phase space we have to use only the existing macro-particles. Secondly, to not affect any distribution functions we can make the down-sampling procedure probabilistic so that for each macro-particle the chance of reducing the weight to zero is balanced by the chance of increasing its weight. 

We thus propose to use the term \textit{agnostic} in relation to the down-sampling that probabilistically changes the weights of macro-particles so that (1) at least one of them receives zero weight, i.e. can be removed, and (2) for each macro-particle the expectation value for the weight is exactly equal to its initial weight. It is clear that in this case all the distribution functions are not changed on average, i.e. the procedure keeps an appropriate chance for all {the features to remain, being affected only in a non-systematic statistical way, which is anyway natural for the approach of statistical representation with macro-particles. The chance of removing all the particles that carry information about some feature is vanishingly small in case there are many such particles. This is what we mean by the term \textit{excessive representation} required for the down-sampling to be applicable in a certain region of the phase space.} Because this procedure implies only the removal of particles it can also be referred to as \textit{thinning out} or just \textit{thinning}.

Let us consider a simple example (referred to as "random" below): for a set of $n > 1$ particles we chose one with equal probability $1/n$ and remove it, while the weight of each other macro-particle in the set is increased by factor $n/\left(n – 1\right)$. Although this procedure obviously satisfies the outlined principle, it conserves neither total weight nor any other quantities that may be of importance. This naturally brings us to the question whether it is possible to conserves several quantities and if so how to arrange such an agnostic conservative down-sampling. We construct the solution in the next section.

\section{The description of algorithm}

Because of the outlined properties, the algorithm can be applied to any subset of macro-particles, such as for all macro-particles in a certain region or a cell of the grid, as well as for macro-particles in a revealed dense cluster. {Note, however, that in order to make a proper use of the method through preserving physical quantities locally we need to select subsets of macro-particles that are close to each other. As it has been rigorously demonstrated the spatial closeness of macro-particles in the subsets is also central for minimizing the resampling-related error according to the Lipschitz metric~\cite{rjasanow.jcp.1996} as well as the Sobolev norm~\cite{rjasanow.jcp.1998}.} 

{The procedure of down-sampling can be applied independently for all the selected subsets.} We therefor formulate the problem for a set of $n$ macro-particles representing the same specie of real particles. For the sake of shortness from hereafter we omit "macro-" when referring to macro-particles, wherever it is not confusing.

Suppose before applying the method the $i$-th particle has the statistical weight $w_i^p$ (hereafter the superscript $p$ denotes "prior"). The set of particles can then be described by an $n$-dimensional vector $\textbf{w}^p = \left(w^p_1, w^p_2, ... w^p_n\right)$. 
Applying the method implies determining the vector $\textbf{w}^a$ ($a$ denotes "after") that has only positive components and at least one component equal to zero. Each particle, that corresponds to zero component, is then removed while others are assigned with new weights that are equal to the corresponding components of $\textbf{w}^a$.

In this notation, the scalar product of vector $\textbf{w}$ and vector $\textbf{1} = \left(1, ... 1\right)$ is equal to the total statistical weight of all the particles within the set $\textbf{w}$:

\begin{equation}\label{n_cons}
W = \left(\textbf{w}, \textbf{1} \right) = \sum_{i = 1}^{n} w_i.
\end{equation}

The requirement that the method conserves the total number of real particles is equivalent to the requirement, that this scalar product remains the same before and after applying the method, i.e. $\left(\textbf{w}^a, \textbf{1} \right) = \left(\textbf{w}^p, \textbf{1} \right)$. 

{Using such vector notation, the conservation of certain scalar products can be used to formulate the fact that the method preserves arbitrary additive quantities that are linear with respect to particles' weights, including conservation laws, contributions to the grid quantities and central moments of particle distribution in the set. In terms of macroscopic treatment, the preserved physical quantities must be linear with respect to particle density in the phase space, but can be nonlinear with respect to the phase-space coordinates.} For example, for the energy conservation the components of the vector, that forms the scalar product, should be equal to the energy of real particles represented by the macro-particles. {Another example: when performing down-sampling of particles in a given cell or dense cluster, the preservation of the grid values for the current requires constructing, for each affected node, the vector from the relative contributions of each particle to the value at this node in accordance with the used form factor (whatever it is). In both cases the nonlinearity in the dependency of contributions on the phase-space coordinates does not interfere with further analysis because we intend to construct the resulting set out of the particles present in the initial set.} 

We assume that we have in total $m$ restrictions of {described} type including the weight conservation (\ref{n_cons}), and we thus need to conserve $m$ scalar products:
\begin{equation}\label{const}
\left(\textbf{w}, \textbf{e}^j \right) = \sum_{i = 1}^{n} e^j_i w_i  = E, \: \: j = 1, ... \: m.
\end{equation}
Here each component of the $j$-th vector $\textbf{e}^j = \left(e^j_1, e^j_2, ... e^j_n\right)$ is defined as the corresponding characteristic/contribution of the related particle.

Our goal can now be formulated. We need to determine several possible outcomes $\textbf{w}^a_k$ and the probabilities $p_k$ of choosing them so that (a) each of $\textbf{w}^a_k$ has only non-negative components and at least one zero component, (b) each of $\textbf{w}^a_k$ satisfies all the constraints (\ref{const}) and (c) for each particle the expectation value of weight is equal its initial weight:
\begin{equation}\label{agn}
\left<\textbf{w}^a\right> =\sum_{k \: \in \: \textrm{\small outcomes}} \textbf{w}^a_k p_k = \textbf{w}^p.
\end{equation}

Let us first consider an arbitrary non-zero vector $\textbf{e}$ and the corresponding constraint $\left(\textbf{w}^p, \textbf{e}\right) = \left(\textbf{w}^a, \textbf{e}\right)$. A vector $\textbf{v} \neq 0$, that is perpendicular to $\textbf{e}$ (i.e. $\left(\textbf{v}, \textbf{e}\right) = 0$), we will call \textit{balanced} relative to constraint $\textbf{e}$. This is because in the context of our problem it satisfies the following property: the result of adding this vector multiplied by any number to $\textbf{w}^p$ satisfies the constraint given by $\textbf{e}$. Clearly, for any given constraint any linear combination of balanced vectors is also a balanced vector.

\begin{figure}
	\centering
	\includegraphics[width=1.0\columnwidth]{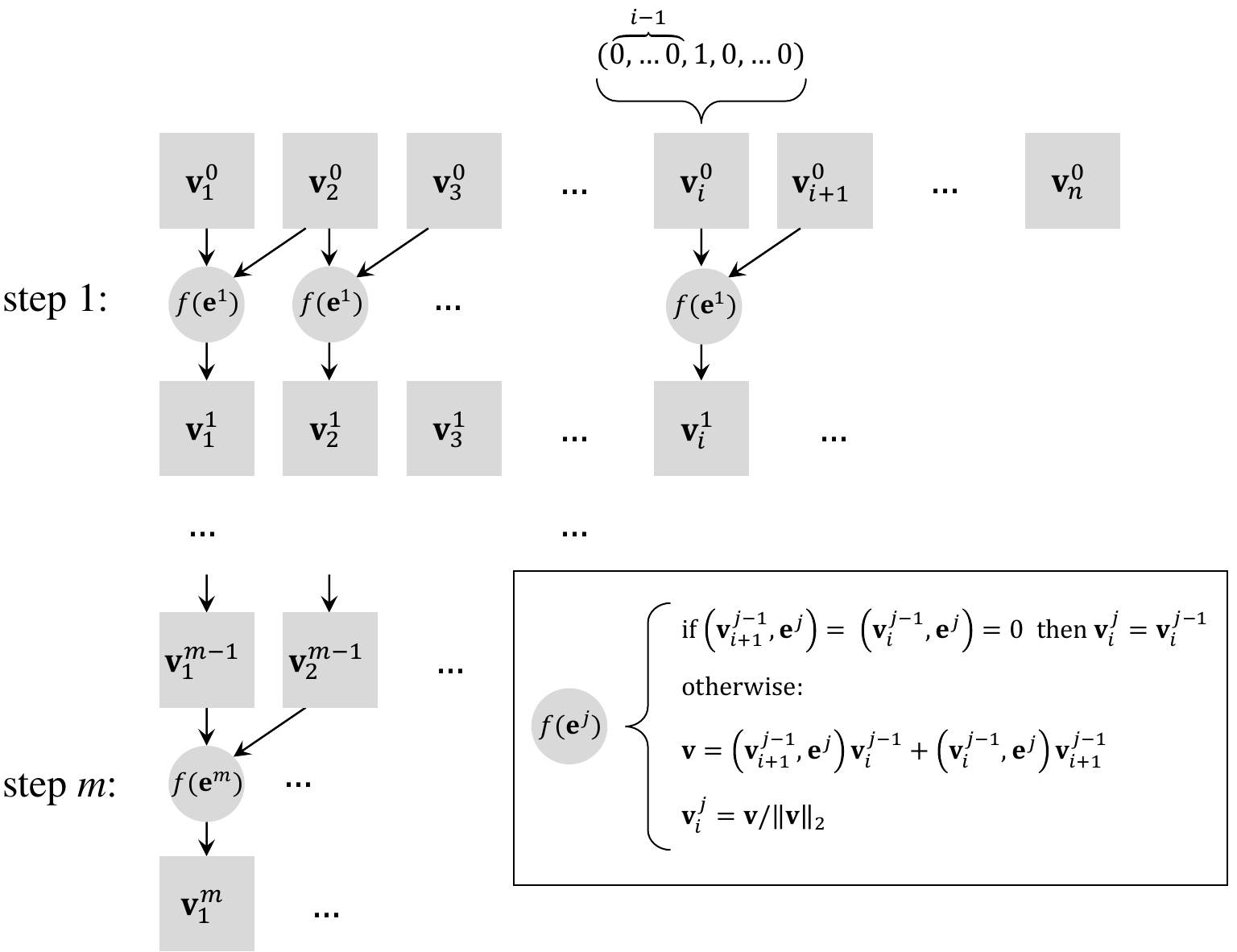}
	\caption{{Schematic illustration of the procedure used for obtaining the vector that is balanced with respect to all the constraints. The steps are presented from the top down (see the description in the text). The pseudo code in the insert details the implementation of the operations $f(\textbf{e}^j)$ on the vectors shown with gray squares.}}
	\label{scheme}
\end{figure}

Next, we note that out of two arbitrary non-collinear vectors $\textbf{a}$ and $\textbf{b}$ we can always construct a non-zero vector that is balanced relative to any given constraint $\textbf{e}$. Indeed, in case either $\left(\textbf{a}, \textbf{e}\right) \neq 0$ or $\left(\textbf{b}, \textbf{e}\right) \neq 0$ we can construct such a vector by
\begin{equation}
\textbf{c} = \left(\textbf{a}, \textbf{e}\right)\textbf{b} - \left(\textbf{b}, \textbf{e}\right)\textbf{a},
\label{perp}
\end{equation}
and if $\left(\textbf{a}, \textbf{e}\right) = \left(\textbf{b}, \textbf{e}\right) = 0$ we can take any linear combination of $\textbf{a}$ and $\textbf{b}$ {(in our implementation we simply take one of the vectors as will be further detailed in fig.~\ref{scheme})}.

{Now we can start constructing the solution to our problem by taking a set of vectors  $\textbf{v}_i$, which spans the space of all possible variations of weights, and then reducing it $m$ times so that the spanned space is restricted sequentially by each of the constraints. In practice we allocate a single set and transform it doing sequentially $m$ steps, but here we denote the state with superscript for better readability. We start from the set of $n$ vectors defined as follows: $i$-th vector $\textbf{v}^0_i$ has $i$-th component equal to 1 and all other components equal to zero.} At the first step we construct $n-1$ vectors  $\textbf{v}^1_i$ that are balanced relative to the first constraint $\textbf{e}^1$. We here simply use the pairs of neighbors, i.e. construct the balanced vector $\textbf{v}^1_i$ out of $\textbf{v}^0_i$ and $\textbf{v}^0_{i+1}$ ($i = 1, ... n-1$). We then repeat this operation sequentially for all the constraints, i.e. for $j = 2, ... m$ we construct vectors $\textbf{v}^j_i$ out of  $\textbf{v}^{j-1}_i$ and $\textbf{v}^{j-1}_{i+1}$ ($i = 1, ... n-j$). At each step the vectors are constructed as linear combinations and thus appear balanced relative to all previously considered constraints, i.e. $\textbf{v}^j_i$ is balanced relative to all $\textbf{e}^k$, $k \le j$. {The described procedure is illustrated in fig.~\ref{scheme}. Note that in practice at each step we normalize the result of eq.~(\ref{perp}) to prevent running into the limitations of numerical arithmetic.}

Now we see that if $n > m$ we can do all the steps and obtain the vector $\textbf{v}^b$ that is balanced relative to all the constraints (exceptions are discussed below). If there are more than one such vectors, one can simply take $\textbf{v}^b = \textbf{v}^m_0$ {as we do in our implementation} (see the related remark below). The vector
\begin{equation}
\textbf{w} = \textbf{w}^p + \alpha \textbf{v}^b
\end{equation}
fulfills all the constraints (\ref{const}) for any value of $\alpha$. Since $\textbf{v}^b$ is balanced relative to the constraint $\textbf{1}$ (weight conservation), it must have at least one positive and at least one negative component. Thus, by increasing the value of $\alpha$ from zero we can reach the point, for which at least one component of $\textbf{w}$ is zero, while all others are positive. This is achieved with
\begin{equation}
\alpha_+ = \min_{v^b_i < 0}\left\{-\frac{w^p_i}{v^b_i}\right\}.
\label{eq:min}
\end{equation}
We can also make at least one component zero, while keeping others positive, by decreasing $\alpha$ from zero:
\begin{equation}
\alpha_- = \max_{v^b_i > 0}\left\{-\frac{w^p_i}{v^b_i}\right\}.
\label{eq:max}
\end{equation}
We can now {propose} the solution to our problem based on these two outlined cases: the algorithm chooses $\textbf{w}^a = \textbf{w}^a_+$ with probability $p_+$ and $\textbf{w}^a = \textbf{w}^a_-$ with probability $p_-$, where
\begin{equation}\label{solution}
\begin{array}{lll}
&\textbf{w}^a_+ = \textbf{w}^p + \alpha_+ \textbf{v}^b, & p_+ = \alpha_- /\left( \alpha_- - \alpha_+\right),\\
&\textbf{w}^a_- = \textbf{w}^p + \alpha_- \textbf{v}^b, & p_- = \alpha_+ /\left( \alpha_+ - \alpha_-\right).\\
\end{array}
\end{equation}
According to the construction procedure, both cases fulfill all the constraints and in both cases at least one component is zero, while all others are positive. The probabilities $p_+$ and $p_-$ are chosen so that each component remains unchanged on average:

\begin{equation}\label{verification}
\begin{array}{lll}
&\left<\textbf{w}^a\right> = \left(\textbf{w}^p + \alpha_+ \textbf{v}^b\right)p_+ + \left(\textbf{w}^p + \alpha_- \textbf{v}^b\right)p_- \\
& = \textbf{w}^p \left(p_+ + p_- \right) + \displaystyle{\frac{\alpha_+ \textbf{v}^b\left(-\alpha_-\right) + \alpha_- \textbf{v}^b \alpha_+}{\alpha_+ - \alpha_-}} = \textbf{w}^p.
\end{array}
\end{equation}
This concludes the verification of the proposed solution.

{We can now outline briefly the steps that are essential for the implementation:
\begin{enumerate}
	\item{Select a group of particles for down-sampling (e.g. considers densely populated cells or use clustering method);}
	\item{Calculate the components of vector $\textbf{e}^{j}$ ($j = 1, ... m$) for each of $m$ quantities to be preserved (see examples below);}
	\item{Initialize vectors $\textbf{v}_i^0$ and do $m$ steps to obtained vector $\textbf{v}_1^m$ that is balanced with respect to all the constraints (see fig.~\ref{scheme} and the description in the text);}
	\item{Use eqs.~(\ref{eq:min}) and (\ref{eq:max}) to determine the candidates for removal and to obtain the values of $\alpha_{+}$ and $\alpha_{-}$;}
	\item{Generate a random number to make a choice based on probabilities given in eq.~(\ref{solution});}
	\item{Remove the selected particle (or particles) and adjust the weights of others accordingly (see eq.~(\ref{solution}));}
	\item{To further reduce the number of particles (down to $m$ in the ultimate case) repeat steps 2~-~6 (alternatively one can apply steps 2~-~6 to subsets of $m+1$ particles randomly chosen from the group selected at step 1).}
\end{enumerate}
}
	
To complete the description we need to make few remarks. Firstly, when constructing the vectors $\textbf{v}^k_i$ we can potentially run into a situation when the vectors in the used pair are exactly collinear. In practice these cases must be exceptionally rare and we therefore can just halt the attempt of down-sampling for the given subset of particles in such cases (although, if $n$ is sufficiently larger than $m$, the solution still may exist and it could be possible to find it). Secondly, the choice of pair for the construction is arbitrary, while we still need to involve all the previous vectors to span the entire space of possibilities. Other selection rules (instead of neighbors) may favor certain logic. Thirdly, if $n > m + 1$ we obtain more than one vector at the final step. These vectors span the space of possibilities for the down-sampling. One can design a procedure to select within this space the option that is favorable in some sense. For example, we can try to minimize the difference between the weights. {Note that eqs. (\ref{eq:min}) and (\ref{eq:max}) suggest that the particles with lower weight are likely to restrict the value of $\alpha$, i.e. these particles are likely to either be removed or get an accordingly increased weight. This way or other, the algorithm in the described form naturally tends to get rid of particles with low weight (relative to other weights). That is fortunate because such particles provide a less efficient use of computational resources. Fourthly, the described algorithm guarantees the removal of only one particle, but if $n > m + 1$ we can apply it sequentially several times to reduce the number of particles in the subset to $n = m$.}

Finally, as we mentioned previously, if the algorithm is applied to the macro-particles in a highly populated cell or cluster, we can configure it to preserve local density and other quantities. For example, apart from the total energy and momentum, we can preserve the local contribution to the density, current density, momentum flow or other grid values of importance. If we need to preserve $m$ quantities of this kind, the algorithm can remove at least one macro-particle from any subset of more than $m$ macro-particles. This means that by sequential use we can down-sample any given set of more than $m$ macro-particles to a subset of $m$ macro-particles. For example, if we use cloud-in-cell weighing we can reduce the number of macro-particles in any given cell to 36 (if it is larger initially) preserving the charge and current density in the cell's nodes ($8 \times (1 + 3)$), as well as the total energy (1) and momentum (3). Alternatively, we can preserve few first central moments to reduce the statistical noise.

{It is also worth noting that the required excess of particle number over the number of preserved scalar quantities ($n > m$) indicates an important practical limitation: the method is applicable only for the regions where the number of particles per cell is larger than the number of preserved quantities. This goes in line with the computational necessity and implies that the method makes it possible to cap the number of particles per cell at the number of preserved quantities. That is why for maintaining reasonable computational costs it may be favorable to have as small value of $m$ as possible, i.e. it is important to preserve only those quantities that are crucial for the problem in question.}
	
\section{Validation}

We chose to use a {couple of very simple examples} in order to disentangle the demonstration from any possible applications and physics, where the use of proposed algorithm may be of interest. {In the first example,} we consider a one dimensional density distribution given by the following expression:
\begin{equation}
D(x) = \left\{
\begin{array}{lll}
&2, \: \: &x \in \left(0.25, 0.5\right],\\
&4 - 16\left(x - 0.5\right), \: \: &x \in \left(0.5, 0.75\right],\\
&0, \: \: &x \in \left(-\infty, 0.25\right] \cup \left(0.75, +\infty\right).\\
\end{array}
	\right.
\end{equation}
Note that $\int D(x) dx = 1$. As the prior ensemble we use the result of sampling this distribution with $N = 2 \times 10^{5}$ macro-particles. To enforce {uniformity} of initial representation while allowing for the variation of weight we do the following. We vary the weight $w$ from 0 to twice the average value $2/N$ and place the macro-particles sequentially (moving in the positive $x$ direction) with an appropriate interval $w/D(x)$ between each other.      

We then compare the results of reducing the number of macro-particles in this representation by factor 2 using one of five downs-sampling methods described below. For each method we select randomly a subset of 5 neighboring macro-particles and apply the down-sampling method. We repeat this procedure until the total number of particles becomes less or equal to $N/2 = 10^5$.

As the first method we use "random" down-sampling, which has been described in the introduction. We remove the macro-particle chosen randomly with equal probability 0.2 and increase the weight of others by 1.25. 

As the second method, referred here as "weight-conservative", we use the following procedure. We select a particle with probability proportional to the weights of particles in the subset, i.e. with probability $w^p_i/W$. We do this selection 4 times and then for each particle assign the weight $s W / 4$, where $s$ is the number of times the particle has been selected. The particles, that have not been selected, are removed. One can check that this procedure fulfills the principle of agnostic down-sampling and also conserves the total weight of the subset.

As the third, fourth and fifth method we use the proposed agnostic conservative down-sampling. As the "conservative-1" we refer to the option that preserves the "zeroth" and the first central moments, i.e. the weight and the mean position of the weight:
\begin{equation}
x_{\textrm{mean}} = W^{-1} \sum w_i x_i,
\end{equation}
where $x_i$ is the position of $i$-th particle. As the "conservative-2" and "conservative-3" we refer to the options that preserve all central moments up to variance and skewness, respectively.

\begin{figure}
	\centering
	\includegraphics[width=1.0\columnwidth]{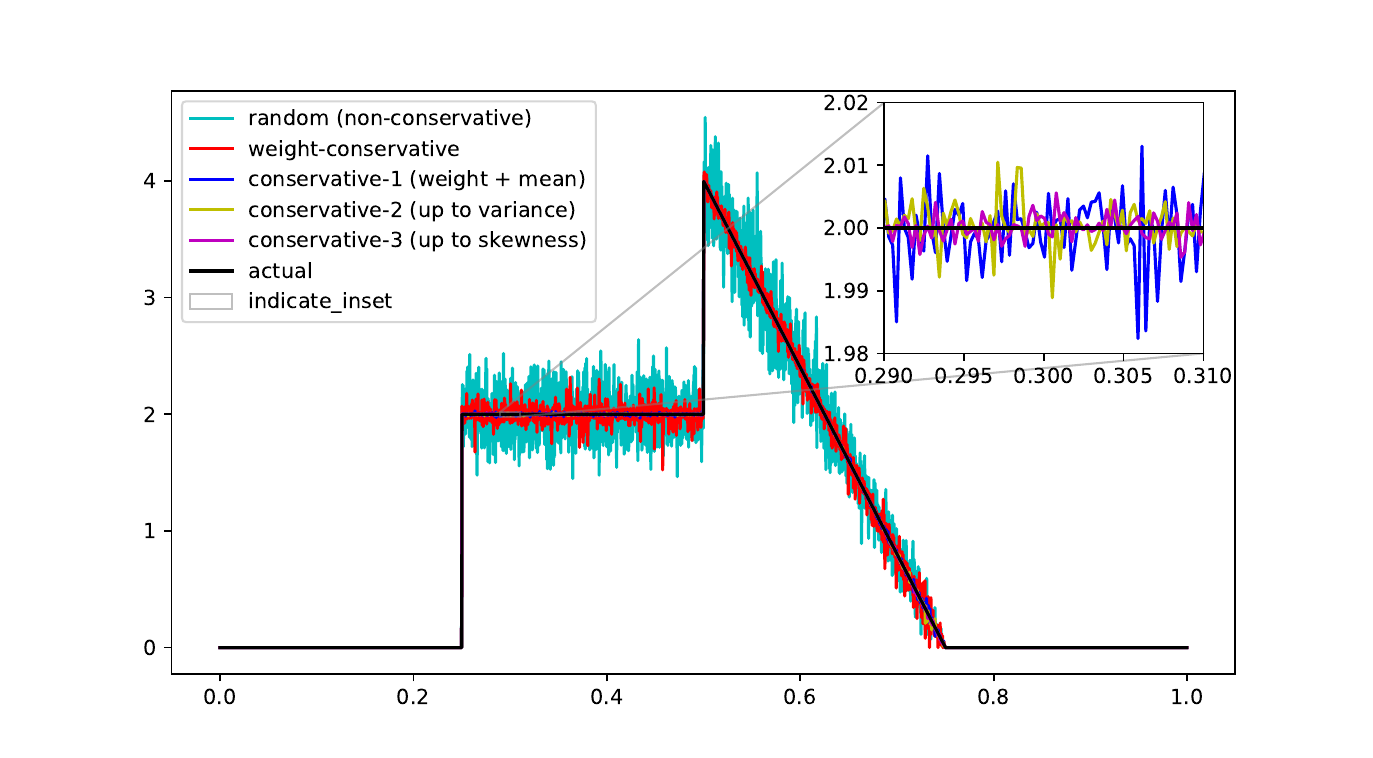}
	\caption{The comparison of different down-sampling methods. {The density of particles is weighted using the cell size equal to 1/4096 and} shown for the cases when we apply one of the following methods: "random" (cyan), "weight-conservative" (red), as well as the proposed method configured to preserve weight and mean coordinate (blue), all the central moments up to variance (yellow) and all the central moments up to skewness (violet). In each case the initial distribution (black) represented by $2 \times 10^5$ macro-particles (with random weights but adjusted spacing) is down-sampled to $10^5$ macro-particles.}
	\label{comp}
\end{figure}

{In fig.~\ref{comp} we show the density of particles obtained at the grid with 4096 cells using the nearest point weighing. First, we can see that in all the cases the down-sampling does not build up any systematic deviations (at any spatial scales) such as flattening that one can expect from, e.g., merging into a center mass point. To test this, we used a distribution with a sharp jump and generated initial particles without using any information about the grid. Note that the procedures of down-sampling are not given with any information about the cell size or any other spatial scales. This observation can be considered as a demonstration of the fact that the used algorithms has the property of being agnostic.}

Form fig.~\ref{comp} we can also see that the constraints imposed on the down-sampling suppress the random variations of density relative to the initial state, while the macro-particle number is the same in all cases. To quantify this, we compare the results in terms of the standard deviation of the density across the cells of the grid. As a reference point, we can estimate that a random distribution of $10^5$ particles among 2048 non-empty cells yields approximately 50 particles per cell and this corresponds to the standard deviation of $\approx 0.08$ for the density. The values of standard deviation of the density distributions obtained in fig.~\ref{comp} are detailed in table \ref{table:1}. One can see that the cases do differ in an expected way: the preservation of larger number of central moments gives lower variance. This provides an indirect confirmation of the capability of the method to preserve several quantities simultaneously. However, to provide a more explicit demonstration of this capability we provide another, rather instructive example.

\begin{table}
	\begin{tabular}{lcc}
		Used method
		& Standard deviation\\
		\hline
		\hline
		random
		& $3.2 \times 10^{-2}$\\
		weight-conservative
		& $7.6 \times 10^{-3}$\\
		conservative-1 (weight + mean)
		& $3.1 \times 10^{-4}$\\
		conservative-2 (up to variance)
		& $2.0 \times 10^{-4}$\\
		conservative-3 (up to skewness)
		& $1.2 \times 10^{-4}$\\
		actual (before down-sampling)
		& $6.9 \times 10^{-5}$\\
		\hline
		\hline
	\end{tabular}
\caption{Standard deviation of the density distribution shown in fig.~\ref{comp}}
\label{table:1}
\end{table}

\begin{figure}
	\centering
	\includegraphics[width=1.0\columnwidth]{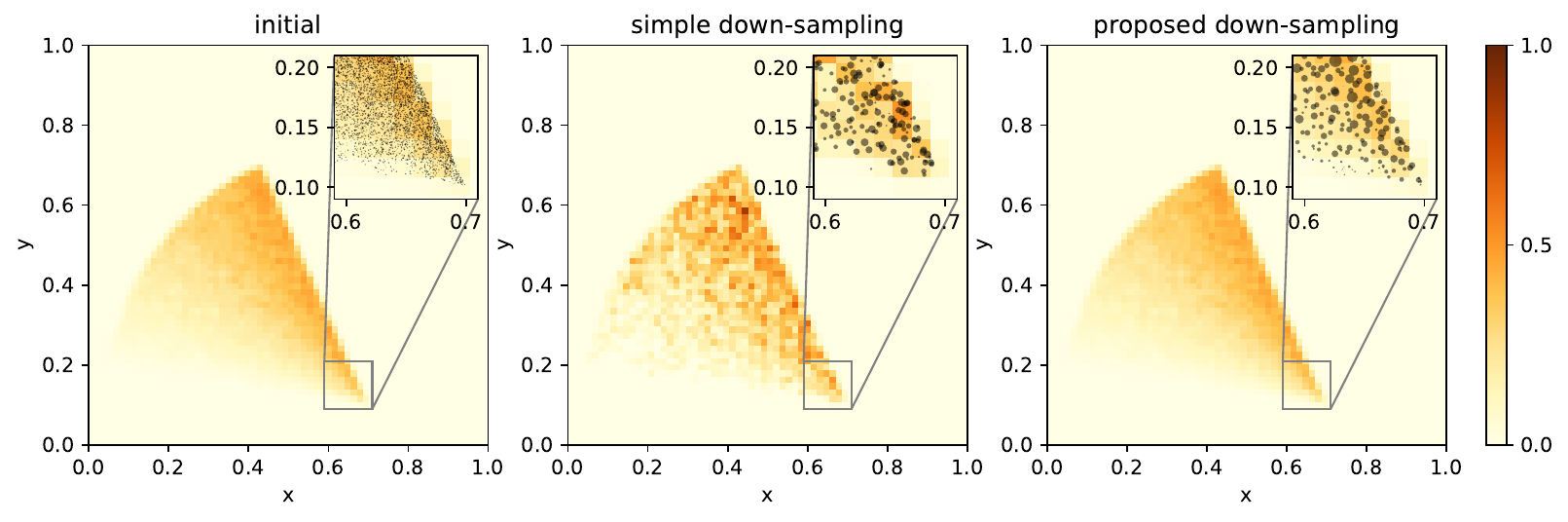}
	\caption{{Demonstration of conservative down-sampling. The CIC-weighed density is shown for the initial distribution (left), for the case of using \textit{simple} down-sampling (center), and for the case of using the proposed method configured to preserve the grid values of density (right). In both cases the final number of macro-particles is approximately 3720, whereas the initial set contains $10^5$ macro-particles. In the inserts the macro-particles are shown with circles, whose areas are proportional to the macro-particles' weights.}}
	\label{comp2d}
\end{figure}

We consider 2D distribution and demonstrate the use of the proposed method for performing down-sampling with conservation of density determined on a grid through cloud-in-cell (CIC) weighing. For the initial set we generate $10^5$ macro-particles with weight $w$ and position $\textbf{r} = (x, y)$ determined by the expressions:
\begin{equation}
\begin{array}{lll}
&{w = 10^{-6} r_1,}\\
&{\textbf{r} = \textbf{c} + r_2^2\left(\cos\left(r_3^{1/2}\right)\textbf{a} + \sin\left(r_3^{1/2}\right)\textbf{b}\right),}\\
\end{array}
\end{equation}
{where $\textbf{c} = (0.7, 0.1)$, $\textbf{a} = (-0.65, 0.1)$, $\textbf{b} = (0.1, 0.65)$, whereas $r_1$, $r_2$ and $r_3$ are uncorrelated random variables that take values from interval $\left[0, 1\right)$ with uniform probability density. The density is calculated on a grid $64\times64$ using CIC weighing. To preserve the gird values of the density, we apply the described method for subsets of particles in between each group of four neighboring nodes, i.e. each subset contains particles that contribute only to the density value at the nodes in a given group (in practice we store particles accordingly). The contribution of particles within a given subset to the density values are considered as four values to be preserved. The coefficients of contributions of each particle form vectors $\textbf{e}^1$, $\textbf{e}^2$,$\textbf{e}^3$ and $\textbf{e}^4$, whereas the total weight is automatically preserved in this case. If the number of particles in a subset is larger than four we can apply the proposed method for conservative down-sampling. To show the ultimate case we use the method to reduce the number of particle to four in each subset with initially larger number of particles. For this we sequentially apply the method to the groups of five particles randomly selected from the remaining ones until we get four particles in the given subset (the random selection is arrange to prevent any possible bias or asymmetry). In such a way we were able to reduce the number of particles to 3716, i.e. the number of particles is reduced by a factor of $f \approx 26.9$. For comparison we also consider down-sampling by the following non-conservative procedure: each particle is either removed with probability $1 - f^{-1}$ or its weight is increased by a factor of $f$. By applying such down-sampling, which we call \textit{simple}, to the initial set, we reduced the number of particles to 3721, which is close to the number of particles after the conservative down-sampling. In fig.~\ref{comp2d} we demonstrate the obtained distributions together with the initial distribution. We can see that the conservative down-sampling indeed preserves the grid values of density. We can also see that the weight of particles in this case varies in a rational way: the regions of higher density are represented by particles with larger weight.}

{Note that we could generate the initial extensive particle set by distributing randomly particles in each cell according to arbitrary density function and then use the described down-sampling to obtain a reduced particle set that perfectly represents the given density according to CIC weighing and does not have any periodic structures or other apparent types of data correlation. This can be useful for arranging a "quiet start" for simulations with PIC method and other statistical methods (see, e.g., \cite{tysanner.ijnmf.2005}).}
	
{From the presented examples we can conclude that the method indeed works in the describe form and provides the opportunity to preserve several quantities simultaneously. According to previous studies, this possibility was considered desirable for various problems of particular interest and we thus hope that the described method will be helpful. The particular choice of quantities to be conserved and the achievable benefits for particular numerical studies should be considered using more problem-motivated characteristic of effectiveness. Some results are presented in Ref.~\cite{muraviev.arxiv.2020}.}

\section{Conclusions}

In this paper we reported on finding a method for performing agnostic conservative down-sampling. The term conservative means that the method can be configured to preserve any number of quantities, such as the total charge, energy and momentum of the given subset of macro-particles, as well as their local contribution to the density, current density, momentum flow or other grid values of importance. Alternatively, the method can preserve few first central moments. The term agnostic is introduced as the property that indicates that the down-sampling procedure changes the weights probabilistically so that they remain unchanged on average. This gives an appropriate chance for keeping {all the meaningful features} in the sampled distribution independently of their scales. The procedure therefore does not require any prior knowledge about the minimal scales of the modeled process in the phase space. We clarified why such an approach may be of interest. In particular, the method may be useful when modelling multi-scale physics or when revealing new phenomena that have yet unknown scales. 

\section*{Acknowledgments}

The research was supported by the Russian Foundation for Basic Research (Projects No. 15-37-21015 and 18-47-520001) and by the Swedish Research Council (Grant No. 2017-05148). {The author would like to thank Valentin Volokitin for useful discussions.} 

\bibliography{literature}

\end{document}